# A Hierarchical Approach to Stability Assessment of Large Scale Interconnected Networks

Thanh Long Vu, *Member, IEEE* and Konstantin Turitsyn, *Member, IEEE*

*Abstract*— Interconnected networks describe the dynamics of important systems in a wide range such as biological systems and electrical power grids. Some important features of these systems were successfully studied and understood through simplified model of linear interconnection of linear subsystems, where provably global properties, e.g. global convergence to a specific state, usually hold true. However, in severely disturbed conditions many of those systems exhibit strongly nonlinear behaviour. Particularly, multiple equilibrium points may coexist and make the dynamical behavior of the system difficult to predict. Aiming at understanding the fragility of interconnected systems, we will provide a hierarchical framework to assess the metastability and resilience of such systems. This framework is based on independently characterizing stability of individual subsystems when they are uncoupled from the network, and then enforcing the diagonal dominance property on a structure matrix capturing the subsystems stability and the input-to-output gains of interconnection network. Since the subsystems are usually of low order and the structure matrix has size equal to the number of subsystems, this framework is easy to implement and thus scalable to large scale interconnected systems. Possible application of this framework in assessing stability of microgrids will be discussed at the end of this paper.

## I. Introduction

Large scale, complex, interconnected system models capture the dynamics of multiple physical, biological, and social structures such as Internet, power grids, brain, and economical systems, etc. In recent years, interconnected systems have been widely investigated to study various dynamical properties of complex systems in biology [1], social sciences [2], physics [3], computer science [4], and engineering [5]. It is well recognized that severe disturbance may expose the fragility of such systems and even lead to catastrophic failure, such as power outages, nervous disorder, and economic collapses. Therefore, it is essential to understand the interconnected systems' stability and resilience, i.e. the ability of the system to withstand some probable disturbance and return to stable operating conditions.

Many works have been then devoted to investigate the dynamical properties of interconnected systems, especially those described by linear interconnection of linear subsystems. For this type of networks, stability analysis and control design are extensively investigated in control community, mainly in the context of consensus in multi-agent systems [6]–[8]. A fortunate feature of these simplified linear models

Thanh Long Vu and Konstantin Turitsyn are with the Department of Mechanical Engineering, Massachusetts Institute of Technology, Cambridge, MA, 02139 USA, e-mail: {longvu, turitsyn}@mit.edu.

is that global stability of the interconnected system can be guaranteed.

It is worth to note that in extreme conditions, many interconnected systems manifest strongly nonlinear behaviour. Remarkably, multiple equilibrium points may coexist, draw a complicated contraction landscape, and make the dynamical behavior of the system difficult to predict. A typical example is the transient dynamics of power grids following a large disturbance, such as line tripping, short circuit, etc. The most simple but practically acceptable model to describe the grid post-fault dynamics is the so-called swing equations:

$$m_k \ddot{\delta}_k + d_k \dot{\delta}_k = P_k - \sum_{\{k,j\} \in \mathcal{E}} B_{kj} \sin(\delta_k - \delta_j). \quad (1)$$

This model captures the dynamics of the generators' rotor angle $\delta_k$ and its angular velocity $\dot{\delta}_k$ after the grid is disturbed. Here, the inherently nonlinear sinusoid functions $B_{kj} \sin(\delta_k - \delta_j)$ representing the electrical power flows between generators in the network $\mathcal{E}$ cannot be ignored. These nonlinear functions result in a system possessing multiple equilibria with their own region of attraction. As such, the transient stability of power grids, after the fault is cleared, can only obtained locally, i.e. when the fault-cleared state stays in a neighborhood of the equilibrium point. Therefore, to understand the fragility of complex structures in response to extreme events, it is imperative to establish a powerful tool capable of assessing such metastability of nonlinear interconnected systems. Unfortunately, there are not many sufficient tools to characterize stability of nonlinear interconnected systems where multiple equilibrium points coexist.

Another main challenge for the stability assessment of large scale interconnected systems is the computational complexity. In the context of power grids, there are thousands to millions of dynamically interacting components, all of which contribute to the stability of the overall grid. This extreme scale causes the grid stability assessment problem a computationally expensive task. One of the efficient approaches, with acceptable computational complexity, is based the hierarchical framework in which the interconnected network is analyzed through a combination of subsystems' stability assessment and network connection assessment. Multiple works utilizing this framework were presented, such as small gain theorem [9], passivity-based approach [10], [11], and contraction analysis of interconnected systems [12]. Again, these works were only devoted to interconnected systems possessing global stability.

In this paper, we aim to alleviate the aforementioned drawback of the global-stability-based approaches, while still targeting at the reduction of computational complexity. Inspired by the works [10]–[12] and connective stability analysis [13], we will present a hierarchical approach to analyze the stability of nonlinear interconnected systems that don't possess global stability. This approach is based on two steps: (i) characterizing the stability of each nonlinear subsystem when it is unperturbed from the network; (ii) estimating the coupling gains in the interconnection network. We prove that if the structure matrix, which captures the stability feature of both individual subsystems and the interconnection gains, is diagonally dominant, then there always exists a linear combination of the subsystems' Lyapunov functions to assessing stability of the whole interconnected systems. As such, the assessment process mostly reduces to establishing Lyapunov functions for the subsystems, which is a computationally tractable task since the subsystems are usually of low order. Therefore, the proposed hierarchical approach is scalable to assessing stability of large scale interconnected systems.

On the other hand, to reduce the possible conservativeness, we exploit the idea of the recently introduced Lyapunov Functions Family method [14] to construction of inner approximations of the attraction region. The principle of this method is to provide stability certificates by constructing a family of Lyapunov functions, and then find the best suited function in the family for given initial states. Accordingly, we present a novel adaptation algorithm capable of finding a suitable Lyapunov function after a finite number of steps. As the last contribution, we show that the proposed approach can be applicable to analysis of the robust stability of interconnected networks with uncertain links and to assessment of system resilience.

The structure of this paper is as follows. In Section II we introduce the model of a linear interconnection of nonlinear subsystems, which lacks of global stability and naturally gives raise to the need of tool for assessing its metastability. On top of this model, we formulate the stability assessment problem. In Section III the main results of the paper are presented where we explicitly construct the family of Lyapunov functions and inscribe the corresponding regions of attraction estimate. Also, we show how to adapt this Lyapunov function family to given initial states and how to robustly assess the stability of the system when the interconnection gains are uncertain. Finally, we present the simulation results in Section IV and conclude the paper in Section V.

## II. SYSTEM MODEL AND PROBLEM FORMULATION

To study the dynamic stability of nonlinear interconnected systems, this paper considers a network interconnecting multiple homogeneous nonlinear subsystems, each of which is described by the following equations:

$$\dot{x}_k = A_k x_k + f_k(x_k) + B_k u_k \quad (2)$$
$$y_k = C_k x_k, k = 1, .., n,$$

where, $x_k \in \mathbf{R}^{n_k}, u_k \in \mathbf{R}$, and $y_k \in \mathbf{R}$ are the state, input, and output of the $k^{th}$ subsystems. Here, $A_k$ is a matrix capturing the linear dynamics of the $k^{th}$ subsystem. By introducing slack variables, many systems in practice can be described by polynomial differential equations, in which case $f_k(x_k)$ characterizes the high order dynamics of the $k^{th}$ subsystem. Note that the analysis in this paper is also applicable to subsystems with multiple inputs and multiple outputs. Let $u = [u_1 \ ... \ u_n]^\top$ and $y = [y_1 \ ... \ y_n]^\top$. The linear network describes the linear relation from the set of network inputs $y = \{y_1, ..., y_n\}$ to the set of network outputs $u = \{u_1, ..., u_n\}$:

$$u = Ly. \quad (3)$$

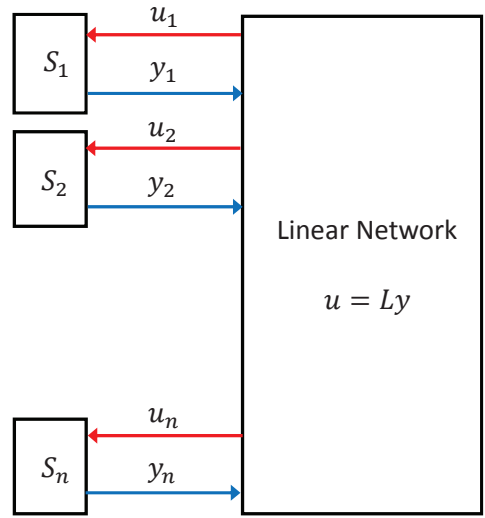

Fig. 1. Linear interconnected networks of homogeneous nonlinear systems as described in (2)-(3)

Here, $L = [L_{kj}]_{k,j=1}^n$ is the matrix characterizing the interconnection in the network: $L_{kj} = 0$ if there is no directed coupling from the the signal $y_j$ to signal $u_k$, and $L_{kj} \neq 0$ if there is a directed coupling from the the signal $y_j$ to signal $u_k$. Substituting (3) into (2), we obtain:

$$\dot{x}_k = A_k x_k + f_k(x_k) + B_k \sum_{j \in \mathcal{N}_k} L_{kj} C_j x_j, k = 1, ..., n \quad (4)$$

where $\mathcal{N}_k$ is the set of indexes of signal $y_j$ coupled with the signal $u_k$. In normal condition, the interconnected system operates at some equilibrium points, which are solutions of the steady state equation:

$$A_k x_k + f_k(x_k) + B_k \sum_{j \in \mathcal{N}_k} L_{kj} C_j x_j = 0, k = 1, ..., n. \quad (5)$$

Since the functions $f_k(x_k)$ are nonlinear functions, there may exist multiple solutions of equation (5), each of which is an equilibrium point of the system. Therefore, the global stability is never obtained for such kind of systems. Particularly, each equilibrium point has its own region of attraction, i.e. the set of initial states from which the system will converge to that equilibrium point.

In this paper, we consider the case that the system is working at a equilibrium point and then disturbed by some disruption, e.g. losing a coupling link. After some time, the failed coupling is fixed and the system returns back to its original topology, but the state of the network deviates from the equilibrium point. We are concerned if the post-fault interconnected system will return back to the original equilibrium point, i.e. if the system withstand the disruption. Formally, the problem considered in this paper is formulated as follows.

> **Stability Assessment:** *Given the post-fault state $x_1(0), ..., x_n(0)$, assess if the network will converge from the state $x_1(0), ..., x_n(0)$ to the equilibrium point $x^* = [x_1^*, ..., x_n^*]$, i.e. if the state $x_1(0), ..., x_n(0)$ stays inside the region of attraction of the stable equilibrium point $x^* = [x_1^*, ..., x_n^*]$.*

To address this problem, we assume that for each unperturbed subsystem (i.e. it is uncoupled from the network) the equilibrium point $x^*$ is locally stable. Mathematically, this condition can be expressed by the following assumptions.

*Assumption 1:* The matrices $A_k$ are Hurwitz, i.e. there exist positive definite matrices $P_k$ and positive numbers $\lambda_k$ such that

$$\lambda_{\max}(A_k^\top P_k + P_k A_k) \leq -\lambda_k, k = 1, ..., n, \quad (6)$$

where $\lambda_{\max}(A)$ denotes the maximum eigenvalue of the symmetric matrix $A$.

*Assumption 2:* There exist positive numbers $d_k, k = 1, ..., n$, such that the higher order dynamics $f_k(x_k)$ is dominated by the linear dynamics in the ball of radius $d_k$ surrounding the equilibrium point $x_k^*$, i.e. there exist positive numbers $\mu_k > 0, k = 1, ..., n$ such that

$$|(x_k - x_k^*)^\top P_k(f_k(x_k) - f_k(x_k^*))| \leq \mu_k \|x_k - x_k^*\|^2, \quad (7)$$

whenever $x_k$ stays in the ball $\|x_k - x_k^*\| \leq d_k$. In this paper, $\|.\|$ denotes the 2-norm of vectors and induced 2-norm of matrices.

## III. Lyapunov Functions Family for Metastability Analysis

### A. Lyapunov Functions Family

To address the stability assessment problem formulated in the previous section we will use a sequence of techniques originating from nonlinear control theory. The procedure here is similar to the connective stability analysis in the work of Šiljak [13], but we aim to analyze the local stability of interconnected systems where the global stability can be never obtained. First of all we rewrite the dynamic equations (4) such that the equilibrium point is shifted to the origin, making it suitable for application of classical stability analysis. Substituting equation (5) into (4), we obtain

$$\dot{x}_k = A_k(x_k - x_k^*) + f_k(x_k) - f_k(x_k^*) + \sum_{j \in \mathcal{N}_k} B_k L_{kj} C_j(x_j - x_j^*), k = 1, ..., n \quad (8)$$

For this network, we propose to use the following family of Lyapunov functions

$$V(x) = \sum_{k=1}^n c_k (x_k - x_k^*)^\top P_k (x_k - x_k^*) \quad (9)$$

where $c_k > 0, k = 1, ..., n$ are constant parameters characterizing the Lyapunov function family. From (8) we have the derivative of the Lyapunov function $V(x)$ along (8) as

$$\dot{V}(x) = \sum_{k=1}^n c_k [\dot{x}_k^\top P_k(x_k - x_k^*) + (x_k - x_k^*)^\top P_k \dot{x}_k]$$
$$= \sum_{k=1}^n c_k (x_k - x_k^*)^\top (A_k^\top P_k + P_k A_k)(x_k - x_k^*)$$
$$+ 2\sum_{k=1}^n c_k (x_k - x_k^*)^\top P_k (f_k(x_k) - f_k(x_k^*))$$
$$+ 2\sum_{k=1, j=1}^n c_k (x_k - x_k^*)^\top P_k B_k L_{kj} C_j(x_j - x_j^*)$$
$$\quad (10)$$

From Assumption 1 and Assumption 2, for any $\|x_k - x_k^*\| \leq d_k, k = 1, ..., n$ we have

$$\dot{V}(x) \leq -\sum_{k=1}^n c_k (\lambda_k - 2\mu_k) \|x_k - x_k^*\|^2$$
$$+ 2\sum_{k=1, j=1}^n c_k (x_k - x_k^*)^\top P_k B_k L_{kj} C_j(x_j - x_j^*)$$
$$\quad (11)$$

Note that $(x_k - x_k^*)^\top P_k B_k L_{kj} C_j (x_j - x_j^*) \leq \|P_k B_k L_{kj} C_j\|.\|(x_k - x_k^*)\|.\|(x_j - x_j^*)\|$. As such, from (11) we have

$$\dot{V}(x) \leq -\sum_{k=1}^n c_k (\lambda_k - 2\mu_k) \|x_k - x_k^*\|^2$$
$$+ 2\sum_{k=1, j=1}^n c_k \|P_k B_k L_{kj} C_j\|.\|x_k - x_k^*\|.\|x_j - x_j^*\| \quad (12)$$

Let $\alpha_{kj} = 2\|P_k B_k L_{kj} C_j\|$ if there is a coupling from $y_j$ to $u_k$, and $\alpha_{kj} = 0$ otherwise. We denote

$$\tilde{x} = [\|x_k - x_k^*\|, ..., \|x_k - x_k^*\|]^\top$$
$$C = \text{diag}(c_1, ..., c_n), \quad (13)$$

and consider the following structure matrix capturing the properties of both individual subsystems and the interconnection network:

$$M = \begin{bmatrix} -\lambda_1 + 2\mu_1 & \alpha_{12} & ... & \alpha_{1n} \\ \alpha_{21} & -\lambda_2 + 2\mu_2 & ... & \alpha_{2n} \\ & & ... & \\ \alpha_{n1} & ... & \alpha_{n(n-1)} & -\lambda_n + 2\mu_n \end{bmatrix}$$

Then, from (12) we conclude that $\dot{V}(x) \leq 0.5 \tilde{x}^\top (M^\top C + CM) \tilde{x}$, for all $\|x_k - x_k^*\| \leq d_k, k = 1, ..., n$. Therefore, we have the following main result of this paper:

*Theorem 1:* Assume that there exists a positive, diagonal matrix $C$ such that the following LMI is satisfied

$$M^\top C + CM < 0 \quad (14)$$

Then, the Lyapunov function $V(x) = \sum_{k=1}^{n} c_k (x_k - x_k^*)^\top P_k (x_k - x_k^*)$ is decreasing whenever $\|x_k - x_k^*\| \leq d_k, \forall k = 1, ..., n$.

Establishing the Lyapunov function $V(x)$ is based on solving the LMI (14) for the given structure matrix $M$ which captures the features of both subsystems stability and interconnection network. This LMI can be formulated as a convex optimization problem. In addition, since the structure matrix $M$ has the size equal to the number of subsystems, which is not very large, this LMI can be quickly solved by the advanced Semidefinite programming solvers (e.g. [15], [16]). Therefore, the proposed approach significantly reduces the computational complexity of the stability assessment process for large scale interconnected systems, in which most of computational load is spent on constructing Lyapunov functions (characterized by positive definite matrices $P_k$) for the subsystems of low orders.

An easy-to-check condition for the existence of the matrix $C$ is given in the following theorem.

*Theorem 2:* Assume that the structure matrix $M$ is diagonally dominant, i.e. $-\lambda_k + 2\mu_k + \sum_{j=1, j\neq k}^{n} \alpha_{kj} < 0$ for all $k = 1, ..., n$. Then, there exists a positive, diagonal matrix $C$ satisfying the LMI (14)

*Proof:* See Appendix.

### B. Constructions of Invariant Sets

The existence of a positive, diagonal matrix $C$ as in Theorem 1 ensures the decreasing of the Lyapunov function $V(x)$ in the set $\mathcal{P} = \{x : \|x_k - x_k^*\| \leq d_k, \forall k = 1, ..., n\}$. There, however, may exist the case when the initial state lies inside $\mathcal{P}$, but after some time periods, the system trajectory escapes from the set $\mathcal{P}$. Then the Lyapunov function $V(x)$ is no longer decreasing, and the system trajectory may tend to undesired region. In order to ensure that the system will not escape the set $\mathcal{P}$ during transient dynamics we will add one condition to restrict the set of initial states inside $\mathcal{P}$.

Formally, we define the minimization of the function $V(x)$ over boundary of $\mathcal{P}$:

$$V_{\min} = \min_{x \in \partial \mathcal{P}} V(x) \quad (15)$$

Due to the independence of $x_k$ from other state $x_j, j \neq k$, it holds that

$$V_{\min} = \min_{k=1,...,n} \min_{\|x_k - x_k^*\| = d_k} c_k (x_k - x_k^*)^\top P_k (x_k - x_k^*).$$

The corresponding invariant set is defined as:

$$\mathcal{R} = \{x \in \mathcal{P} : V(x) < V_{\min}\}. \quad (16)$$

The decay property of Lyapunov function in the set $\mathcal{P}$ ensures that the system trajectory cannot escape $\mathcal{R}$. Therefore, staring from any initial state inside the set $\mathcal{R}$, the system will only evolve inside $\mathcal{R}$ and eventually converge to the desired stable equilibrium point $x^*$ due to the negative definiteness of $\dot{V}(x)$ inside the set $\mathcal{R}$. As such, to check if a given initial state $x(0)$ will lead to stable operating condition $x^*$, we only need to check if $x(0) \in \mathcal{R}$, i.e. if $x(0) \in \mathcal{P}$ and $V(x(0)) < V_{\min}$.

### C. Adaptation of Lyapunov Functions to Initial States

The family of Lyapunov functions characterized by the diagonal matrix $C$ satisfying the LMI (14) allow us to find a Lyapunov function that is best suited for a given initial state $x_0 \in \mathcal{P}$ or family of initial states. In the following, we apply the stability certificate in Theorem 1 and propose a simple algorithm for the adaptation of Lyapunov functions to a given initial state $x_0$.

Let $\epsilon$ be a positive constant.

- *Step 1:* Find $C^{(1)}$ by solving the LMI (14). Calculate $V^{(1)}(x_0)$ and $V^{(1)}_{\min}$ where

$$V^{(1)}(x) = \sum_{k=1}^{n} c_k^{(1)} (x_k - x_k^*)^\top P_k (x_k - x_k^*).$$

- *Step n:* If $x_0 \notin \mathcal{R}(C^{(n-1)})$, then find $C^{(n)}$ by solving the following LMIs:

$$M^\top C^{(n)} + C^{(n)} M < 0, \quad (17)$$

$$\sum_{k=1}^{n} c_k^{(n)} (x_k(0) - x_k^*)^\top P_k (x_k(0) - x_k^*) \leq V^{(n-1)}_{\min} - \epsilon \quad (18)$$

Suppose that at Step $n$, we still have $x_0 \notin \mathcal{R}(C^{(n)})$, i.e., $V^{(n)}(x_0) \geq V^{(n)}_{\min}, \forall i = 1, .., n$. Then,

$$V^{(n)}_{\min} \leq V^{(n)}(x_0) = \sum_{k=1}^{n} c_k^{(n)} (x_k(0) - x_k^*)^\top P_k (x_k(0) - x_k^*)$$

$$\leq V^{(n-1)}_{\min} - \epsilon \leq ... \leq V^{(1)}_{\min} - (n-1)\epsilon. \quad (19)$$

Since $V^{(n)}(x)$ is lower bounded, this algorithm will terminate after a finite number of the steps. There are two alternatives exit then. If $V^{(n)}(x_0) < V^{(n)}_{\min}$, then the Lyapunov function is identified. Otherwise, the value of $\epsilon$ is reduced by a factor of 2 until a valid Lyapunov function is found. Therefore, whenever the stability certificate of the given initial condition exists, this algorithm possibly finds it after a finite number of iterations.

### D. Robust Stability Assessment

In practice, there may exist uncertainties in the system dynamics, e.g. the coupling gains of the linear network are unknown. However, we assume that these gains are bounded in the means that:

$$|L_{kj}| \leq \bar{l}_{kj}, \forall k, j \quad (20)$$

Let $\bar{\alpha}_{kj} = 2\bar{l}_{kj} \|P_k B_k\| \|C_j\|$ if there is a coupling from $y_j$ to $u_k$, and $\alpha_{kj} = 0$ otherwise. Similar to Section III.A, we can

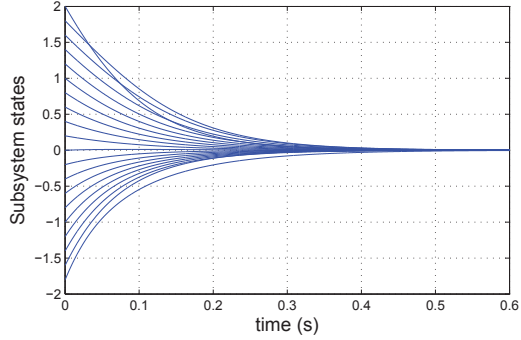

Fig. 2. Convergence of the subsystem states $x_k, k = 1, ..., 20$, from the initial state to the equilibrium point

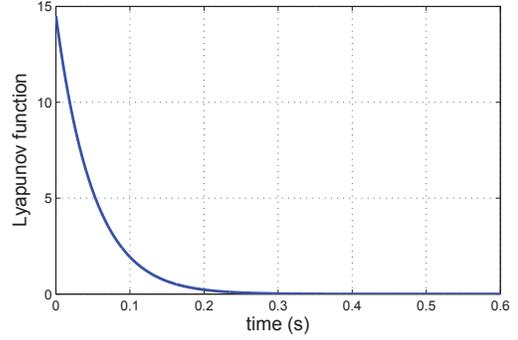

Fig. 3. Convergence of the Lyapunov function $V(x)$

prove that $\dot{V}(x) \leq 0.5\tilde{x}^\top(\bar{M}^\top C + C\bar{M})\tilde{x}$, for all $x \in \mathcal{P}$, where $\bar{M}$ is the robust structure matrix:

$$\bar{M} = \begin{bmatrix} -\lambda_1 + 2\mu_1 & \bar{\alpha}_{12} & ... & \bar{\alpha}_{1n} \\ \bar{\alpha}_{21} & -\lambda_2 + 2\mu_2 & ... & \bar{\alpha}_{2n} \\ & & ... & \\ \bar{\alpha}_{n1} & ... & \bar{\alpha}_{n(n-1)} & -\lambda_n + 2\mu_n \end{bmatrix}.$$

Therefore, if the robust structure matrix $\bar{M}$ is diagonally dominant, then from Theorem 2 we can find a positive, diagonal matrix $C$ such that the matrix $(\bar{M}^\top C + C\bar{M})$ is negative definite. Then we have the same stability analysis as in Sections III.B and III.C. It is worth to note that this robust stability certificate is applicable to some networks with nonlinear couplings, such as power grids where the power flows are nonlinear but linearly bounded.

*E. Resilience Analysis*

The stability certificate presented in the previous sections allows us to analyze the resilience of interconnected networks when there are some possible losings of the interconnection link. Indeed, we can see that if the subsystems are sufficiently stable and the gains of interconnection network are not very large such that the robust structure matrix $\bar{M}$ is diagonally dominant, then we can still find the Lyapunoc function $V(x)$ characterized by the diagonal matrix $C$ even when there are some permanently lost links. Therefore, in this case the system is still capable of withstanding the disturbance of permanently losing links and returning to its stable operating condition. In case when the subsystems are not sufficiently stable, resilience analysis can be performed in term of estimating the time period for losing links [17].

IV. NUMERICAL ILLUSTRATION

To illustrate the effectiveness of the proposed approach we consider an artificial time-varying interconnected network of the following subsystems

$$\dot{x}_k = -10x_k + a_k x_k^2 + u_k \quad (21)$$
$$y_k = x_k, k = 1, ..., 20 \quad (22)$$

We are interested to assess the stability of some initial state with respect to the equilibrium point $x_k^* = 0, k = 1, ..., 20$. The gain of high order dynamics satisfies $|a_k| \leq 1$. The interconnection network couples each subsystem with its two around subsystems and constitutes a loop. Particularly, the network is time-varying and described by

$$u = L(t)y$$

where the interconnection matrix $L(t)$ is given by

$$\begin{bmatrix} 0 & L_{1,2}(t) & 0 & ... & L_{1,20}(t) \\ L_{2,1}(t) & 0 & L_{2,3}(t) & ... & 0 \\ & & ... & & \\ 0 & ... & L_{19,18}(t) & 0 & L_{19,20}(t) \\ L_{20,1}(t) & 0 & ... & L_{20,19}(t) & 0 \end{bmatrix} \quad (23)$$

Assume that the time-varying coupling strength satisfy: $|L_{ij}(t)| \leq \bar{l}_{ij} < 2$ for all $i, j$.

Consider the Lyapunov function $V(x) = \sum_{k=1}^{20} c_k \frac{x_k^2}{2}$. Let $d_k = 6$. Then the robust structure matrix $\bar{M}$ in (21) can be calculated as

$$\begin{bmatrix} -10 + 6|a_1| & \bar{l}_{1,2} & 0 & ... & \bar{l}_{1,20} \\ \bar{l}_{2,1} & -10 + 6|a_2| & \bar{l}_{2,3} & ... & 0 \\ & & ... & & \\ \bar{l}_{20,1} & 0 & ... & \bar{l}_{20,19} & -10 + 6|a_{20}| \end{bmatrix}$$

Since $|a_k| \leq 1$ and $\bar{l}_{kj} < 2$ for all $k, j$, the matrix $\bar{M}$ is diagonally dominant. According to Theorem 2 there always exists a positive, diagonal matrix $C$ such that $\bar{M}^\top C + C\bar{M}$ is negative definite. Then, the corresponding Lyapunov function $V(x) = \sum_{k=1}^{20} c_k \frac{x_k^2}{2}$ can be used to assess the stability of the system. The set $\mathcal{P}$ now becomes $\{x : |x_k| \leq 6\}$ and the invariant set is $\mathcal{R} = \{|x_k| \leq 6 : V(x) < V_{\min}\}$ where $V_{\min} = \min_{k=1,...,20} 18c_k$.

For numerical illustration, we take the following data set: $a_k = 0.9$, for $k = 1, ..., 10$, $a_k = k/20$, for $k = 11, ..., 20$. The interconnection matrix is given by $L_{k,k+1} = 1.9 \sin x_k, k = 1, ..., 20$, and $L_{k+1,k} = -1.8 \cos x_k, k = 1, ..., 20$. Hence, we can take $\bar{l}_{k,k+1} = 1.9, \bar{l}_{k+1,k} = 1.8$. This data set ensures that the matrix $\bar{M}$ is diagonally dominant and the stability certificate can be performed. We take the initial state $x_k(0) = 0.2(k-10), k = 1, ..., 20$. Using Matlab CVX software for the adaptation algorithm presented in Section III.C, after 2 steps we can find a Lyapunov function $V(x)$ which certifies the convergence of the network from this initial state to the concerned equilibrium point $x_k^* = 0, k = 1, ..., 20$. For this Lyapunov function we have $V_{\min} = 17.9277$ and $V(x(0)) = 14.4952$. Figures 2 and 3 confirm the convergence of the time-varying interconnected network from the initial state $x(0)$ to the equilibrium point $x_k^* = 0, k = 1, ..., 20$.

## V. Conclusions and Path Forward

This paper has presented a scalable framework for assessing the transient stability of interconnection of nonlinear subsystems, where many equilibrium points may coexist and the global stability is never obtained. The proposed method was based on two independent steps: (i) characterizing the stability of each nonlinear subsystem when it is uncoupled from the network; (ii) estimating the input-to-output gains in the interconnection network. The stability of the interconnected nonlinear systems was then assessed through the establishment of a linear combination of Lyapunov functions of the subsystems. This combination was shown to exist for systems with diagonally dominant structure matrix. The stability condition on the interconnection network was derived in the linear matrix inequality (LMI) form with linear matrix size equal to the number of subsystems.

By construction, the proposed approach is scalable to large interconnected systems, such as power grids. We also derived an explicit inner approximation of the stability region of the equilibrium point, and showed how the Lyapunov functions can be adapted to certify transient stability for any given initial state. Finally, we discussed how the proposed approach can be applied to stability assessment of uncertain interconnected networks.

In the future we plan to explore possible strategies of reducing the overall conservativeness of the proposed certificates. We also plan to extend the framework to nonlinear descriptor systems commonly used for modeling power systems and develop more systematic approaches for estimation of subsystem gains based on modern algebraic techniques [18], [19].

## VI. Appendix: Proof of Theorem 2

Given the $\infty$−norm on the Euclidean space and its induced matrix norm $\|A\|_\infty$, the associated matrix measure of $A$ is defined (see e.g. [20]) as the one-sided directional derivative of the matrix norm in the direction of the $A$ evaluated at the identity, i.e.

$$\mu_\infty(A) = \lim_{h \downarrow 0} \frac{\|I + hA\|_\infty - 1}{h} \quad (24)$$

In can be proved that for $A = [a_{ij}]_{i,j=1}^n$ we have $\mu_\infty(A) = \max_{i=1,...,n}(a_{ii} + \sum_{j=1,...,n, j \neq i} |a_{ij}|)$. Therefore, the diagonal dominance property of $M$ implies that the $\infty$−norm induced matrix measure of the matrix $M$ is negative. Note that the $\infty$−norm induced matrix measure of the matrix $M$ is always larger than or equal to $\max_{1 \leq k \leq n} \mathbf{Re}\lambda_k(M)$ [21], where $\lambda_k(M), k = 1, ..., n$, are eigenvalues of $M$. Hence, we conclude that $\mathbf{Re}\lambda_k(M) < 0$ for all $k = 1, ..., n$. Therefore, the matrix $(-M)$ is an M-matrix and thus, there exists a positive, diagonal matrix $C$ such that $(M^\top C + CM)$ is negative definite [22].